\documentclass[preprint,showpacs,aps]{revtex4}
\usepackage{graphicx}
\usepackage{color} 
\usepackage{dcolumn}
\usepackage{bm}
\usepackage{amssymb}

\preprint {IMSc/2011/10/14}
\begin{document}
\title{Group velocity of neutrino waves} 
\author{D. Indumathi$^1$, Romesh K Kaul$^1$, M.V.N. Murthy$^1$
and G. Rajasekaran$^{1,2}$}

\affiliation
{$^1$ The Institute of Mathematical Sciences, Chennai 600 113, India.\\
$^2$ Chennai Mathematical Institute, Siruseri 603 103, India.\\
}
\date{\today}

\begin{abstract} We follow up on the analysis of Mecozzi and Bellini 
(arXiv:1110:1253v1) where they showed, in principle, the possibility of 
superluminal propagation of neutrinos, as indicated by the recent OPERA 
result. We refine the analysis by introducing wave packets for the 
superposition of energy eigenstates and discuss the implications of 
their results with realistic values for the mixing and mass parameters 
in a full three neutrino mixing scenario. Our analysis shows the 
possibility of superluminal propagation of neutrino flavour in a very 
narrow range of neutrino parameter space. Simultaneously this reduces
the number of observable events drastically. Therefore, the OPERA result
cannot be explained in this frame-work.

\vspace{4cm}

\begin{quote}
{\sf We dedicate this paper to the memory of Raju Raghavan
who has made fundamental contributions in the area of neutrino physics,
and who passed away while we were writing this paper.}
\end{quote}

\vspace{4cm}

\end{abstract}

\pacs{14.60.Pq, 
96.40.Tv, 
95.55.Vj 
}

\maketitle

\section{Introduction} 

The recent announcement of the OPERA result \cite{opera} indicating 
possible superluminal propagation of neutrinos has excited considerable 
interest. Various aspects of the experiment, the analysis of the data 
and their interpretation, must be subjected to a thorough examination 
since the result has important repercussions on fundamental physics. 
Furthermore, independent confirmation or refutation by other experiments 
is absolutely essential.

If the result stands, one must first see whether it can be understood 
within the usual framework of physics before giving up cherished 
notions such as Lorentz invariance. In fact, this is possible, in 
principle, as was shown by Mecozzi and Bellini \cite{mb}. By 
considering the interference between the different mass eigenstates of 
the neutrinos they showed that superluminal propagation is possible. We 
extend their analysis by explicitly including the effect of the finite
width of wave packets and provide a numerical estimate of the effect for
realistic neutrino parameters with three generation mixing and matter
effects taken into account.

Our numbers indicate that there is a very narrow region in the allowed
parameter space with three neutrino flavour mixing in which superluminal
propagation is possible in principle. However, the survival probability
for neutrinos with superluminal velocities is almost vanishing rendering
them unobservable in practice.  Furthermore, this depends crucially on the
ratio of the distance of propagation and the energy of the neutrinos. At
distances and energies corresponding to the OPERA experiment, superluminal
propagation is not possible with the present limits imposed by the
neutrino parameter space. If the OPERA result is confirmed it would
require new physics. Therefore, we would like to stress the importance
of further studies along the present lines since the OPERA experiment
has opened up a new window on neutrino physics, which may be called
neutrino optics and which should be pursued by future experiments.

In Sec.~II we outline the group velocity calculation in the wave 
packet formalism. In the context of neutrino oscillation, this formalism
has been discussed in detail in Ref.~\cite{akhmedov}. We derive results
for short and very long base-line propagation of neutrinos, based on which
we present a realistic numerical analysis in the framework of 3 neutrino
mixing in Sec.~III. Other issues related to group velocity measurements
will be discussed in Sec.~IV while Sec.~V concludes the paper with some
remarks on the implications of OPERA-type experiments in future.

\section{Calculation of the group velocity}

The neutrino flavour states $|\nu_\alpha\rangle,~\alpha=e,\mu,\tau$ are 
related to mass eigenstates $|\nu_i\rangle, ~i=1,2,3$ by
\begin{equation}
|\nu_\alpha\rangle=\sum_{i}U_{\alpha i}|\nu_i\rangle,
\label{eq1}
\end{equation}
where $U$ is a unitary matrix. For a neutrino that starts as a flavour 
state $\alpha$ at $t=0$, the state vector at time $t$ is 
\begin{equation}
|\psi(t)\rangle = N\int~dp~ 
\sum_{i}U_{\alpha i}~e^{ipx-iE_it}~e^{-(p-p_0)^2/a^2}|\nu_i\rangle,
\label{eq2}
\end{equation}
where we have superposed the three energy eigenstates with the same momentum 
p and then superposed different $p$ with an amplitude of gaussian form 
$e^{-(p-p_0)^2/a^2}$ to form a wave packet. We have set $\hbar=c=1$.
The normalisation constant is determined by the condition 
$\langle \psi(t)|\psi(t)\rangle =1$.

The probability amplitude for detecting $|\nu_\beta\rangle$ at time $t$ is 
\begin{eqnarray}
A_{\beta\alpha}&=&\langle \nu_\beta|\psi(t)\rangle = N\int dp
\sum_{i}U_{\alpha i}U^*_{\beta i} ~e^{ipx-iE_it}~e^{-(p-p_0)^2/a^2}
\\
&=& N\int dp \sum_{i}C_i
^{\beta\alpha}
~e^{ipx-iE_it} ~e^{-(p-p_0)^2/a^2},
\label{eq3}
\end{eqnarray}
where we have used
$$
\langle \nu_\beta|\nu_i\rangle =U^*_{\beta i};
~~~C_i^{\beta\alpha}
 = U^*_{\beta i}U_{\alpha i}~.
$$
We expand the energy $E$ around the peak $p_0$ of the gaussian and
keep upto the quadratic terms (assuming the width of the wave packet in
momentum space to be small enough): 
\begin{equation}
E \approx E_0 + \frac{dE}{dp}
\left \vert \rule{0pt}{16pt} \right. _{p_0} (p-p_0) +
\frac{1}{2}\frac{d^2E}{dp^2}
\left \vert \rule{0pt}{16pt} \right. _{p_0} (p-p_0)^2
+\cdots~.
\label{eq4}
\end{equation}
With this approximation, the $p$-integration in eq.~(\ref{eq3}) can be done to 
yield the result (absorbing the constant factors into $N$):
\begin{equation}
A_{\beta\alpha}=N\sum_{i}C_i^{\beta\alpha}
~e^{ip_0x-i\widetilde{E}_{0i}t} 
~e^{-(x-x_0^i(t))^2a^2/4};~~~\widetilde{E}_{0i}=E_{0i}+\frac{a^2}{4}
\frac{d^2E_i}{dp^2}
\left \vert \rule{0pt}{16pt} \right. _{p_0}~,
\label{eq5}
\end{equation}
where $x_0^i(t) = dE_i/dp\left\vert_{p_0}\right. t$. Thus we have a
superposition of 3 gaussians in $x$-space with their centres travelling 
with 3 separate group velocities,
\begin{equation}
v^i \equiv \frac{dx_0^i(t)}{dt} = \frac{dE_i}{dp}
\left \vert \rule{0pt}{16pt} \right. _{p_0}~.
\end{equation}
However, if the $x$-space gaussians are broad, that is, if $a$ is small,
then these 3 gaussians will interfere. To study this, let us define
\begin{equation}
\langle x \rangle_{\beta\alpha} =
\frac{\int~x~ |A_{\beta\alpha}|^2 dx}{\int |A_{\beta\alpha}|^2 dx}~.
\label{eq6}
\end{equation}
The integrations are straightforward and the result is
\begin{equation}
\langle x \rangle_{\beta\alpha} =
\frac{\sum_i|C_i^{\beta\alpha}|^2 v_i t+\sum_{i>j} Re(C_i^{\beta\alpha}
C_j^{\beta\alpha *}(v_i+v_j) t
~e^{-i\Delta E_{ij}t})~e^{-(v_i-v_j)^2t^2a^2/8}}
{\sum_i|C_i^{\beta\alpha}|^2+2~\sum_{i>j} Re(C_i^{\beta\alpha} 
C_j^{\beta\alpha*} 
~e^{-i\Delta E_{ij}t})~e^{-(v_i-v_j)^2t^2a^2/8}}~,
\label{eq7}
\end{equation}
where 
$$
\Delta E_{ij}=E_{0i}-E_{0j}+\frac{a^2}{4}\frac{d^2(E_i-E_j)} {dp^2}
\left \vert \rule{0pt}{16pt} \right. _{p_0}~.
$$
Defining the overall-group velocity of neutrinos generated as
$\nu_\alpha$ at time $t=0$ and detected as $\nu_\beta$ at time $t$
as $v_{\beta\alpha}$ as was done by Mecozzi and
Bellini \cite{mb}, we get
\begin{equation}
v_{\beta\alpha}=\frac{\langle x \rangle_{\beta\alpha}}{t} =
\frac{\sum_i|C_i^{\beta\alpha}|^2 v_i +\sum_{i>j} Re(C_i^{\beta\alpha}
C_j^{\beta\alpha *}(v_i+v_j)
~e^{-i\Delta E_{ij}t})~e^{-(v_i-v_j)^2t^2a^2/8}}
{\sum_i|C_i^{\beta\alpha}|^2+2~\sum_{i>j} Re(C_i^{\beta\alpha}
C_j^{\beta\alpha*} 
~e^{-i\Delta E_{ij}t})~e^{-(v_i-v_j)^2t^2a^2/8}}~.
\label{eq8}
\end{equation}
This is the main result of the paper. 

We now consider the special case of two generation mixing with $\mu,\tau$
as the two neutrino flavours.
Substituting $\alpha=\beta=\mu$, and denoting $C_1^{\mu\mu}=\cos^2\theta, 
C_2^{\mu\mu}=
\sin^2\theta$, we have
\begin{equation}
v_{\mu\mu}=
\frac{v_1\cos^4\theta + v_2 \sin^4\theta + (v_1+v_2)
\sin^2\theta\cos^2\theta\,
\cos(\Delta E_{12}t)~e^{-(v_1-v_2)^2t^2a^2/8}}
{\cos^4\theta +\sin^4\theta +2\sin^2\theta\cos^2\theta
\,\cos(\Delta E_{12}t)~e^{-(v_1-v_2)^2t^2a^2/8}} ~.
\label{eq9}
\end{equation}
Let us consider the two extreme limits from the above equation. 
First consider the case when the width $a^{-1}$ of the wave packets
in the $x$-space is large compared to the distance of separation between the 
two centres of the wave packets, that is $a(v_1-v_2)t \ll 1$. In this limit 
eq.~(\ref{eq9}) becomes
\begin{eqnarray}
v_{\mu\mu}&=&
\frac{v_1 \cos^4\theta + v_2 \sin^4\theta + (v_1+v_2)
\sin^2\theta\cos^2\theta \,
\cos(\Delta E_{12}t)}
{\cos^4\theta +\sin^4\theta +2\sin^2\theta\cos^2\theta \,
\cos(\Delta E_{12}t)}\nonumber\\
&=& \frac{1}{2}\left[(v_1+v_2)+(v_1-v_2)\frac{\cos2\theta}
{1-\sin^2 2\theta~\sin^2(\Delta E_{12}t/2)}\right]~.
\label{eq10}
\end{eqnarray}
This agrees with the result of Mecozzi and Bellini\cite{mb}. 

However as the
distance or time of propagation increases, the width of the wave packets
in $x$-space becomes small compared to the distance of separation between 
the two centres of the wave packets, $a(v_1-v_2)t \gg 1$, and we get
\begin{equation}
v_{\mu\mu}=
\frac{v_1 \cos^4\theta +v_2 \sin^4\theta }
{\cos^4\theta +\sin^4\theta} =
\frac{1}{2}\left[(v_1+v_2)+(v_1-v_2)\frac{\cos2\theta}
{1-\frac{1}{2}\sin^2 2\theta}\right]~,
\label{eq:gaussdamp}
\end{equation}
which is the weighted average of the group velocities of the two wave 
packets. Interestingly, the effect of the gaussian suppression is
precisely the same as taking the average over energy and distance in the
denominator of eq.~(\ref{eq10}). This makes sense since, for instance,
at astrophysical distances, the neutrino wave-length is small compared
to the distance of propagation.

Thus the result in eq.~(\ref{eq8}) generalises the result of 
Mecozzi and Bellini to wave packets of finite width. Because
of the approximation made in eq.~(\ref{eq5}), for large $a$
the damping factor $\exp({-(v_1-v_2)^2t^2a^2/8})$ is only approximate,
although its exact replacement also will damp the oscillatory factor
$\cos (\Delta E_{12}t)$.
 
If the calculations are done for waves of infinite spatial extent, the
integrals occurring in the numerator and denominator of eq.~(\ref{eq6})
would be individually divergent, although the final result would turn out
to be finite. However, it is better to do these calculations with wave
packets of finite width $a$, as has been done above, and take the limit
of $a\to 0$ in the end.

We have superposed the 3 mass eigenstates with the same momentum $p$ but
different $E_i$ to form the neutrino state vector in eq.~(\ref{eq2}).
Should one superpose different $p_i$ but the same $E$, or, different
$p_i$ and different $E_i$? This question has been studied in recent
literature \cite{process}. Such possibilities will be included in a
future study of the group velocities of neutrino waves, which is under 
preparation.

For neutrinos whose energy is very large compared to their rest masses, the 
formula in eq.~(\ref{eq9}) may be written in the form
\begin{equation}
v_{\mu\mu}=
\frac{v_1\cos^4\theta + v_2 \sin^4\theta + (v_1+v_2) \sin^2\theta\cos^2\theta 
~\cos y~e^{-(\delta p/p)^2y^2/4}}
{\cos^4\theta +\sin^4\theta +2\sin^2\theta\cos^2\theta
~\cos y~e^{-(\delta p/p)^2y^2/4}}~,
\label{eq12}
\end{equation}
where $y=\Delta m^2L/(2E)$, $\Delta m^2$ and $L$ being the measured
mass-squared difference and base-line distance and $\delta p=a/\sqrt{2}$
the width of the $p$-space wave packet. Thus the exponential damping
factor multiplying the oscillatory factor $\cos y$ is simply $e^{-(\delta
p/p)^2y^2/4}$ where $\delta p/p$ is the fractional uncertainty in
momentum.

We also note that the formula in eq.~(\ref{eq9}) and eq.~(\ref{eq8}) for 
two- and three- generations are valid, to the leading order, for 
propagation through matter at constant density when vacuum values of 
$E_i$ and the mixing angles are replaced by their matter-dependent 
values.

From eq.~(\ref{eq5}), the normalised oscillation probability, that is,
the probability for detecting flavour $\beta$ at time $t$ is
\begin{equation}
P_{\beta\alpha}={\int \vert A_{\beta\alpha} \vert^2 dx}
= \sum_i \vert C_i^{\beta\alpha} \vert^2 + \sum_{i > j} 2 \hbox{Re } 
(C_i^{\beta\alpha} C_j^{\beta\alpha*}
e^{-i\Delta E_{ij} t}) e^{-(\delta p/p)^2 y^2/4}~.
\label{eq:probb}
\end{equation}
This differs from the usual
oscillation formula by the factor $\exp(-(\delta p/p)^2 y^2/4)$ in the
second term. In view of the successful
neutrino oscillation phenomenology achived so far, we will assume that
$a = \sqrt{2} \delta p$ is so small that this factor can be
replaced by unity for all the terrestrial experiments as well as solar
neutrino experiments.

Since the oscillation probability given in eq.~(\ref{eq:probb}) is the
denominator in eq.~(\ref{eq8}), the group velocity can become very large
if the oscillation probability is very small. In fact, it can become
infinite if the oscillation probability at that distance is zero. This
is the origin of superluminal propagation, as our analysis in the next
section will clearly show.

We now come back to the interpretation of $\langle x\rangle _{\beta\alpha}$
defined in eq.(\ref{eq6}) which is the basis of the above formulae.
Note that in the denominator we have $\int dx |A_{\beta\alpha}|^2$ instead
of $\sum_\beta\int dx |A_{\beta\alpha}|^2$ which is unity. Thus 
$\langle x\rangle_{\beta\alpha}$ must be interpreted as the {\it conditional 
measurement} of the position of the neutrino under the condition that only
$\nu_\alpha$ is detected. Here the probability amplitude for 
detecting it as $\nu_\alpha$ itself is regarded as the wave function for 
normalising the expectation value of $x$.  This is to be contrasted with
the usual definition of the expectation value of $x$, independent of
the flavour detected, namely
\begin{equation}
\langle x \rangle_{\alpha} =
\frac{\sum_\beta\int~x~ |A_{\beta\alpha}|^2 dx}
{\sum_\beta\int |A_{\beta\alpha}|^2 dx} = 
{\sum_\beta\int~x~ |A_{\beta\alpha}|^2 dx}~.
\label{uncond}
\end{equation}
We may distinguish this case by calling it the {\it unconditional
measurement} of the expectation value of $x$.

Before we go to the numerical analysis, we make some general remarks. As
already pointed out, the origin of the superluminal propagation is
the vanishing of the oscillation probability $P_{\beta\alpha}$ in the
denominator of eq.~(\ref{eq8}). In other words, superluminal propagation and the
vanishing of the oscillation probability go together. Since the number of
events also vanishes, one has the paradoxical situation of {\em
unobservable superluminal propagation}. Our numerical analysis below is
subject to this criticism.

In the particular case of the OPERA experiment, $P_{\beta\alpha}$
becomes the survival probability $P_{\mu\mu}$. There is no evidence for
the survival probability $P_{\mu\mu}$ in OPERA becoming vanishingly
small. Hence explanation of superluminality through enhancement of the
group velocity in the standard oscillation frame-work as discussed above
is untenable.

Our numerical analysis will be based on the group velocity derived from
the conditional measurement of $\langle x \rangle_{\beta\alpha}$ defined
in eq.~(\ref{eq6}), following Mecozzi and Bellini~\cite{mb}.
Alternatively, one could base the analysis on the group velocity derived
from the unconditional measurement $\langle x\rangle_\alpha$ defined in
eq.~(\ref{uncond}). Since this does not have the vanishing denominator,
it will not lead to superluminal group velocity.

Actually, for the OPERA experiement, both $\langle x
\rangle_{\beta\alpha}$ and $\langle x \rangle_{\alpha}$ will give
essentially the same result for the group velocity since $P_{\mu\mu}$
at OPERA does not deviate very much from unity, as far as is known.

\section{Numerical analysis with three generations}

To obtain a realistic estimate for the group velocity of muon neutrinos, 
and its implications, we discuss the three generation scenario.

As shown in the previous section, the gaussian smearing will have no
effect if $(\delta p/p)^2 \ll 1$. We will work under this assumption
and make estimates for this quantity later, showing that it is indeed
small. As a first approximation we neglect the matter effects.

Following eq.~(\ref{eq8}), we may write the ``group velocity" of the 
superposition that starts as $\nu_\mu$ and is detected as $\nu_\mu$
(which we denote as $v_\mu$ for simplicity) as
\begin{eqnarray}
v_\mu & = & v_2 + S^\mu_{12}(v_1-v_2) +S^\mu_{32} (v_3-v_2), \nonumber
\\
 & \equiv & v_2 + \Delta v_\mu~.
\label{eq:vmu}
\end{eqnarray}
The factors $S_{ij}$ are given as
\begin{equation}
S^\mu_{12} = \frac{1}{P_{\mu\mu}} \, \vert U_{\mu 1}\vert ^2[1-2\vert U_{\mu
2}\vert ^2 \sin^2(\Delta E_{21} t/2) -2 \vert U_{\mu 3}\vert ^2
\sin^2(\Delta E_{31}t/2)]~;
\label{eq:S12}
\end{equation}
\begin{equation}
S^\mu_{32} = \frac{1}{P_{\mu\mu}} \, \vert U_{\mu 3}\vert ^2[1-2\vert U_{\mu
1}\vert ^2 \sin^2(\Delta E_{31} t/2) -2 \vert U_{\mu 2}\vert ^2
\sin^2(\Delta E_{32}t/2)]~,
\label{eq:S32}
\end{equation}
where the denominator, which is simply the survival probability of
$\nu_\mu$, is given by,
\begin{equation}
P_{\mu\mu} = 1-4\sum_{j>i=1}^3|U_{\mu i}|^2 |U_{\mu j}|^2 \sin^2(\Delta
E_{ij}t/2)~.
\label{eq:D}
\end{equation}
Here 
\begin{equation}
\frac{\Delta E_{ij} t}{2}
=\frac{(E_i-E_j) t}{2} \approx 
1.27\frac{\Delta_{ij} (\mbox{eV}^2) L(\mbox{km})}{E(\mbox{GeV})}~,
\label{eq16}
\end{equation}
where $\Delta_{ij}=m^2_i-m^2_j$ and the mixing parameters are given
in the basis where the charged lepton mass matrix is diagonal as
\begin{equation}
U=\left[ \begin{array}{ccc}
         c_{13}c_{12} & c_{13}s_{12} & s_{13}~e^{-i\delta} \\
-c_{23}s_{12}-s_{23}s_{13}c_{12}~e^{i\delta} 
& c_{23}c_{12}-s_{23}s_{13}s_{12}~e^{i\delta} 
& s_{23}c_{13} \\ 
s_{23}s_{12}-c_{23}s_{13}c_{12}~e^{i\delta} 
& -s_{23}c_{12}-c_{23}s_{13}s_{12}~e^{i\delta} 
& c_{23}c_{13} \end{array}\right],
\label{eq:U} 
\end{equation} 
where $c_{ij}=\cos \theta_{ij},~s_{ij}=\cos\theta_{ij}$ and $\delta$
is the CP phase. The coefficients $U_{\mu i}$ correspond to the second
row of the mixing matrix. Note that in the absence of any mixing $v_\mu$
in eq.~(\ref{eq:vmu}) reduces to $v_2$, as it should.

We now present a numerical analysis of the possible values of $v_\mu$ 
with realistic parameter values for the mixing angles and mass squared 
differences. We use typical values/ranges for the mass-squared
differences and neutrino mixing parameters: $\Delta_{21}=7.6 \times
10^{-5}$ eV$^2$, $\vert \Delta_{32} \vert = 2 \times 10^{-3}$ eV$^2$,
$\theta_{12} = 34^\circ$, $36^\circ \le \theta_{23} \le 54^\circ$,
$\theta_{13} \le 10^\circ$ while $\delta_{CP}$ is unknown \cite{par}.  The
sign of $\Delta_{32}$ is not known and there are two possible hierarchies,
$m_3^2 >m_2^2$ known as normal hierarchy (N) and $m_3^2<m_2^2$ known as
inverted hierarchy (I). We consider results for both these cases.

\begin{figure}[tbp]
\includegraphics[width=0.48\textwidth]{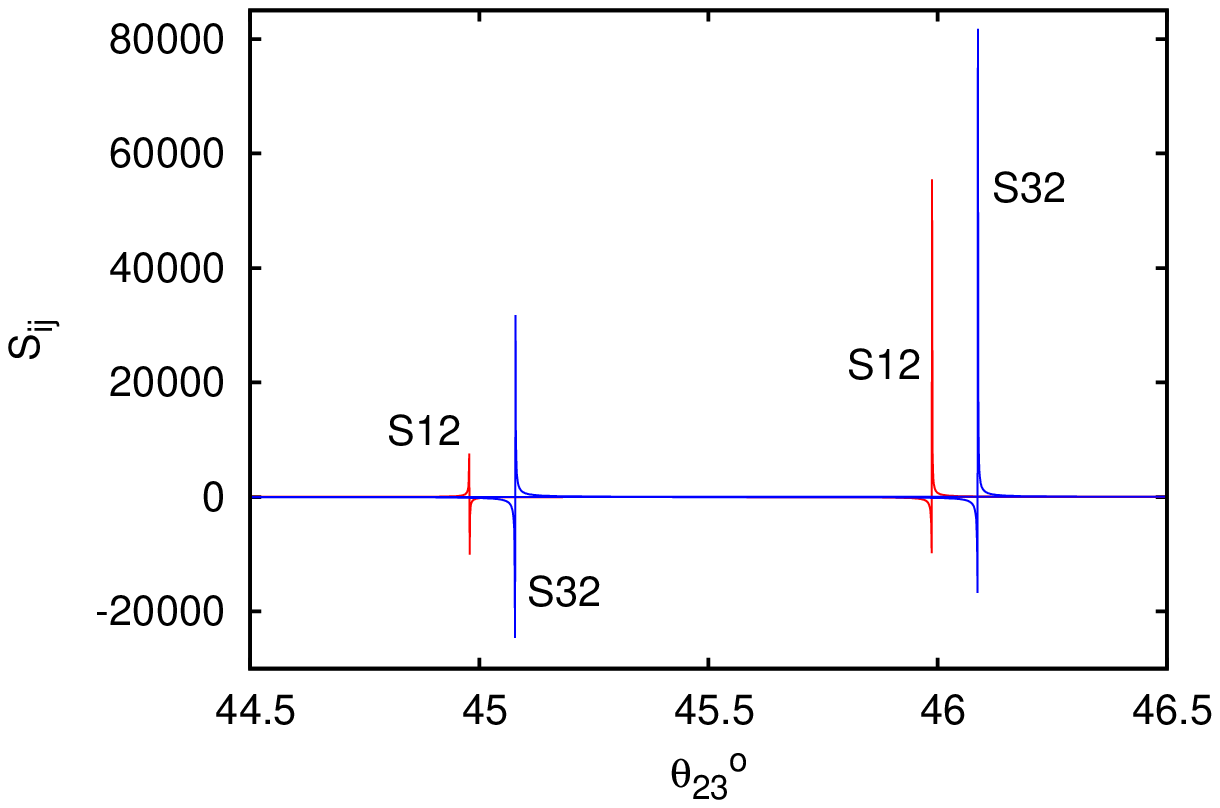}
\includegraphics[width=0.48\textwidth]{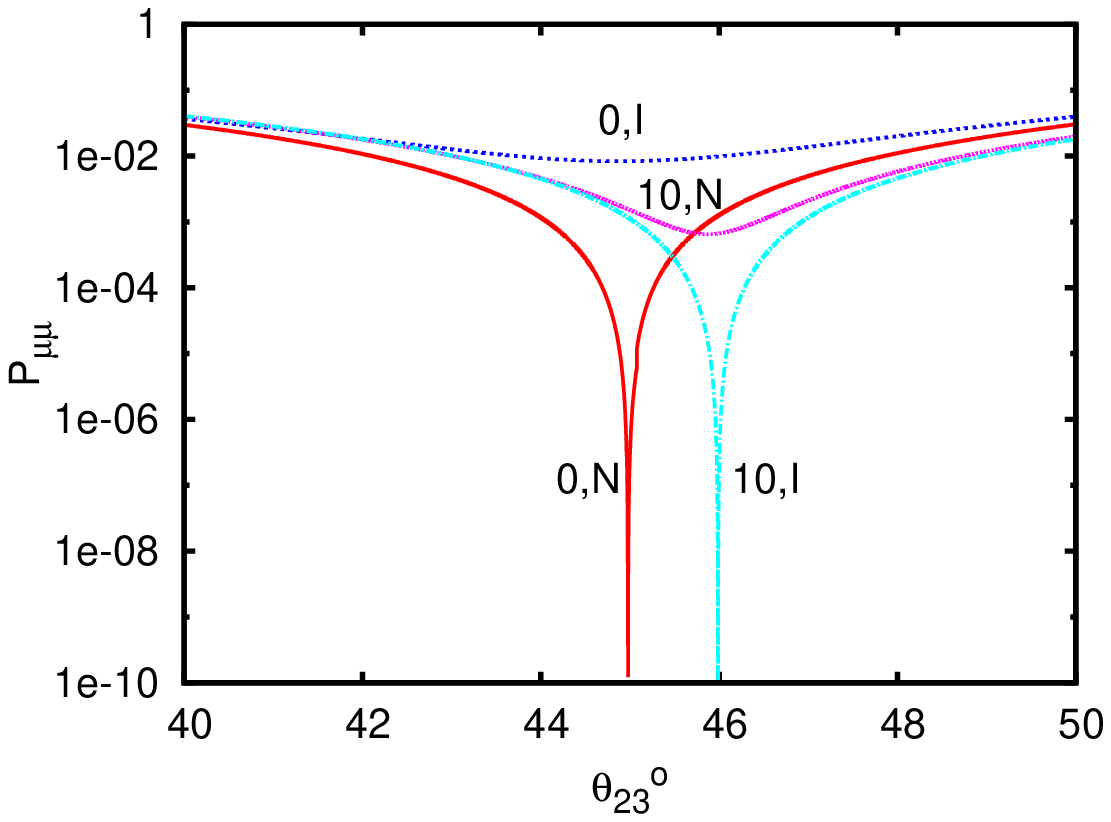}
\caption{(L) The terms $S_{12}$ and $S_{32}$ as
a function of $\theta_{23}$. The values on the left are
for the normal hierarchy solution with $\theta_{13}=0$ while those on
the right are for the inverted hierarchy with $\theta_{13}=10^\circ$. In
both sets, the curves for $S_{32}$ have been offset by
$\theta_{23}=0.1^\circ$ for clarity, else the two curves for $S_{12}$
and $S_{32}$ would overlap each other. In the former case,
$\delta_{CP} = 0$ and $L/E = 611.165$ km/GeV while for the latter,
$\delta_{CP} = 180^\circ$ and $L/E = 604.629$ km/GeV.
(R) The denominator $P_{\mu\mu}$ plotted as a function of $\theta_{23}$
for the same parameters. The dips as $P_{\mu\mu}\to 0$ correspond to
the sharp peaks in $S_{ij}$ on the left.}
\label{fig:Sij}
\end{figure}

Note that the propagation can become superluminal when the $S_{ij}$
are significantly enhanced over unity. Fig.~\ref{fig:Sij} shows the values
of $\theta_{23}$, close to maximal mixing, where this enhancement is
seen, for different mass hierarchies as well as values of $\theta_{13}$.
(Notice therefore that this measurement is sensitive to the mass
ordering in the 2--3 sector which is as yet unknown).

This enhancement of the group velocity occurs for
$L/E~(\hbox{km/GeV}) \sim 600$, well within the range of the OPERA
experiment, viz., $7<L/E (\hbox{ km/GeV}) <730$.

However, it can be seen from the right hand panel of Fig.~\ref{fig:Sij}
that $P_{\mu\mu}$ is very close to zero precisely at these values,
that is, the enhancement of $S_{ij}$ is dominantly due to the vanishing of the
denominator. (The maxima of $S_{ij}$ are slightly offset from the minima of
$P_{\mu\mu}$ because of the dependence of the numerator on the mixing
parameters as well. In fact, when $\theta_{13}=0$, $P_{\mu\mu}$ and
$S_{ij}$ both vanish near $\theta_{23} = 45^\circ$.) This in turn means
that there will be hardly any events that survive at these values,
for this conditional measurement! 

Several remarks are in order:
\begin{itemize}

\item The first is that the mixing parameters as well as $L/E$ need to 
be extremely fine-tuned in order to get this effect. However they are
almost washed out by the suppression of the event rate due to small
survival probability. 

\item The extreme enhancement of $S_{ij}$ over unity is required in
order to obtain superluminal velocities commensurate with the OPERA
observation. This is because the coefficients of $S_{ij}$ which are
$(v_i-v_j)$ in eq.~(\ref{eq:vmu}) behave as $\Delta_{ji}/(2p^2)$
and are small due to the highly suppressing factor of $2p^2$ in the
denominator. It is unlikely that such small excesses will be measurable
by any experiment. The larger the enhancement required, the more finely
tuned are the parameters.

\item Moreover, the fine-tuning in the value of $L/E$ would imply that
the enhancement of the velocity only occurs for a single $E$ value when
the base-line distance $L$ is kept fixed, in contrast to the observed
roughly constant enhancement over a range of $1 \lesssim E \lesssim 100$
GeV as observed in OPERA \cite{opera}.

One way of working around this limitation is to consider the energy
dependence of the matter-dependent contributions. The usual electro-weak
interactions in matter lead to the inclusion of a matter-dependent
potential, $V_{EW}$, that alters the matter dependent mass-squared
differences in a non-trivial way; however, the resulting change is not
large enough to give the required enhancement of the factor of 14 in
the term $\Delta^m_{32} L/(4E)$ at $\langle E \rangle = 17$ GeV.

One possible solution is the inclusion of a new matter interaction
$V_{new}$ with the same energy behaviour, but about an order of magnitude
larger than the usual electro-weak potential. Then, at energies around
$\langle E \rangle = 17$ GeV relevant to the OPERA data, the expression
for the matter-dependent mass-squared differences simplifies to
\begin{equation}
\Delta^m_{ij} \approx \Delta_{ij} + 2E V_{new}~,
\end{equation}
so that the term occurring in $\Delta v_\mu$ becomes
\begin{equation}
\frac{\Delta^m_{ij}L}{4E} \approx \frac{\Delta_{ij}L}{4E} +
\frac{V_{new}L}{2}~.
\end{equation}
The energy dependence of this matter potential
is exactly what is required so that the ratio $\Delta^m_{ij}/E$ is
approximately independent of $E$ for $E \sim 10$s of GeV (but does not
significantly alter the atmospheric neutrino analysis). Such an
energy-independent contribution would remove the fine-tuning in $L/E$
that currently occurs in the expressions for $\Delta v_\mu$ and would
in principle uniformly allow for superluminal propagation of all
velocities relevant to the OPERA analysis. However, while resolving the
fine-tuning in $L/E$, the new matter potential enhances the relevant
terms in a manner identical to the vacuum analysis, viz., through the
vanishing of $P_{\mu\mu}$. Hence, it runs into the same difficulties
with observing the effect as discussed earlier.

\item In spite of what is said above, superluminal group velocity is a
real effect and may be observable in future, as shown in the following
example. Consider the
2-flavour case where we neglect $m_2$ and set $m_3 = \Delta_{32}
\sim 2\times 10^{-3}$ eV$^2$. For smaller energies of the order
of KeVs, observable excesses of $v_\mu$ over unity can be obtained
for a modest value of $P_{\mu\mu} \sim 0.03$. This corresponds
to $\sin^22\theta_{23} = 0.97$, with the ratio $L/E$ tuned so that
$\sin^2(\Delta_{32} L/(4E)) = 1$ (achievable with energies in KeV and
length in meters); this leads to $v_\mu = 1 - \Delta_{32}/(2p^2)\times
(0.5-\cos2\theta_{23}/P_{\mu\mu}) \sim  1 + 6 \, \Delta_{32} /(2p^2)$. For
energy $E$ in KeV we get an enhancement $v_\mu -1 = 6\, 10^{-9}/E^2$;
it may thus be possible to observe the effect since $P_{\mu\mu}$ is not
zero. Of course, there are many experimental problems to be overcome
before such an observation is achieved.

\end{itemize}

\section{Other implications of group velocity measurement}

We add a few remarks on the implication of the OPERA result 
independent of the fact that it is superluminal or subluminal.
Consider a possible measurement of group velocities of electron
and muon neutrinos in a future possible experiment. Following
eq.~(\ref{eq:vmu}), these velocities in vacuum may be written in the form
\begin{eqnarray}
v_e &=& v_1 + S^e_{21}(v_2-v_1) +S^e_{31} (v_3-v_1)~; \nonumber\\
v_\mu &=& v_2 + S^\mu_{12}(v_1-v_2) +S^\mu_{32} (v_3-v_2)~, 
\end{eqnarray}
where $S^\mu_{ij}$ are given in eqs.~(\ref{eq:S12}) and (\ref{eq:S32})
and $S^e_{ij}$ are given by
\begin{equation}
S^e_{21} = \frac{|U_{e 1}|^2[1-2|U_{e 2}|^2 \sin^2(\Delta E_{12} t/2)
-2 |U_{e 3}|^2 \sin^2(\Delta E_{13}t/2)]}
{1-4\sum_{j>i=1}^3|U_{e i}|^2 |U_{e j}|^2 \sin^2(\Delta E_{ij}t/2)}~,
\label{eq18}
\end{equation}
and
\begin{equation}
S^e_{31} = \frac{|U_{e 3}|^2[1-2|U_{e 1}|^2 \sin^2(\Delta E_{13} t/2)
-2 |U_{e 2}|^2 \sin^2(\Delta E_{23}t/2)]}
{1-4\sum_{j>i=1}^3|U_{e i}|^2 |U_{e j}|^2 \sin^2(\Delta E_{ij}t/2)}~.
\label{eq19}
\end{equation}
Simultaneous measurement of these two velocities immediately gives
information on the ordering of masses $m_2$ and $m_3$ since we already
know that $m_2>m_1$ from the solution to the solar neutrino problem.
Unlike earlier proposed solutions \cite{dm}, this does not require
matter effects to resolve the issue. Even a short base-line experiment
with neutrino factories with muon storage rings may help resolve the
hierarchy issue. Note that the $v_e$ and $v_\mu$ could refer to either
neutrinos or antineutrinos. The caveat is that the effect is magnified
only for a given set of parameter values including $L/E$.

\section{Conclusions and some remarks}

To summarise, we have considered the superposition of 3 energy (mass)
eigenstates with the same momentum $p$, but with different energies $E_i$
to form a neutrino flavour state. We have included the effect of the finite
width of wave packets of the same width and considered its effect on
the group velocity of the neutrino flavour state.

When this is applied to the propagation of $\nu_\mu$ for distances of the
order of hundreds of kms and energies corresponding to the observation
of muon neutrinos in the OPERA experiment we find the effect is very small.
Effectively the finite width can be ignored as is done in Ref.~\cite{mb}.

However the effect of width increases with distance. For astronomical 
distances the effect of finite width is to reduce the group velocity to
a weighted average of the velocities of the individual energy eigenstates.
This explains why, even if the OPERA result is correct, there is no 
contradiction with the absence of superluminal propagation of neutrinos 
from supernova SN1987a.

Our analysis shows that superluminal propagation of neutrinos occurs
whenever the oscillation probability corresponding to that measurement
vanishes. For instance, muon neutrinos are dominantly observed at OPERA;
these will exhibit superluminal behaviour precisely when the survival
probability $P_{\mu\mu}$ vanishes. While $P_{\mu\mu}$ does not vanish
for the parameter range of the OPERA measurement, in principle, this
feature makes superluminal neutrinos effectively unobservable through
conditional measurement as explained in section 2.

Finally we discuss the implication of the OPERA result irrespective
of superluminality or otherwise. The neutrino mass
ordering or hierarchy is not fully known. The present understanding is
that this requires separate measurement of oscillations of neutrinos
and anti-neutrinos in the presence of matter. The OPERA measurement of
neutrino flavour ``velocity" has added a new way of precision measurement
of neutrino parameters and may have significance in the context of
neutrino mass hierarchy (provided the parameters lie in an extremely
narrow part of the known range of neutrino parameters). A detailed
analysis including the effect of matter will be presented elsewhere.

After most of the work reported here was completed, we came across
Ref.\cite{berry} which overlaps with parts of the present paper. We
thank Tim R. Morris for bringing to our attention his paper
Ref.~\cite{morris} in which similar ideas have been discussed. In
particular, the extreme fine-tuning of parameters required to produce
velocities of the order observed in OPERA and the difficulty that the
effect is seen only at fixed energy was pointed out in this paper. This
paper points out the existence of multiple peaks, that would give
rise to a strong energy dependence, in addition.

We are grateful to G. Date for many discussions, clarifications and
comments.  We also thank N.D. Hari Dass, S. Kalyana Rama, D. Sahoo,
N. Sinha and R. Sinha for many discussions.

\end{document}